\documentclass[hyper]{JHEP3}
\usepackage{epsfig}
 \usepackage{amsmath}
 \usepackage{amssymb}
 \usepackage{amsfonts}
 \usepackage{bm}

\newcommand{\be}{\begin{equation}}
\newcommand{\ee}{\end{equation}}
\newcommand{\bea}{\begin{eqnarray}}
\newcommand{\eea}{\end{eqnarray}}
\newcommand{\nn}{\nonumber}

\title{K\"{a}hler quantization of $H^*(\mathbb{T}^2,\mathbb{R})$ and modular forms}
\author{Farhang Loran\\
  Department of  Physics, Isfahan University of Technology (IUT)\\
  Isfahan,  Iran\\
  E-mail: \email{loran@cc.iut.ac.ir}}

 \abstract{K\"{a}hler quantization of $H^1(\mathbb{T}^2,\mathbb{R})$ is studied. It is shown that this theory corresponds to
 a fermionic $\sigma$-model targeting a noncommutative space. By solving the complex-structure moduli independence
 conditions,  the quantum background independent wave function is obtained. We study the transformation of
 the wave function under modular transformation. It is shown that
 the transformation rule is characteristic to the operator ordering. Similar results are obtained for
  K\"{a}hler quantization of $H^2(\mathbb{T}^2,\mathbb{R})$.}
 \keywords{Topological Field Theories, Topological Strings}
\begin{document}
 \section{Introduction}\label{Int}
 Recently in \cite{Aganagic} it is shown that the topological string
 partition function in real polarization is a holomorphic function
 but not a modular form in the usual sense. The partition functions of the topological B-model on a Calabi-Yau threefold satisfy the holomorphic
 anomaly equation \cite{BCOV}
 which can be obtained by K\"{a}hler quantization of $H^3({\rm CY}_3,\mathbb{R})$
 \cite{Witten,Verlinde,Kahler}. In this paper, we study K\"{a}hler quantization of $H^1(\mathbb{T}^2,\mathbb{R})$.
 This theory is equivalent
 to a fermionic $\sigma$-model $\mathbb{R}\to{\cal P}_2$ where ${\cal P}_2=\mathbb{R}^2/{\rm SL}(2,\mathbb{Z})$ is the two dimensional
  space of periods of $\Omega\in H^{1}(\mathbb{T}^2,\mathbb{R})$.
  The quantum background independent wave function can be obtained
 by solving the complex-structure moduli independence
 conditions \cite{Kahler}. We show that in real polarization, the wave function  is a
  {\em quasi-modular} form of weight one\footnote{By definition a modular form of weight $k$, is a holomorphic
  function $f(\tau)$ which transforms as
  \be
  f({\bm\tau})\to \left(c{\bm\tau}+d\right)^kf({\bm\tau}),
  \ee
  under modular transformation.
  Here we define a quasi-modular form of weight $k$ by transformation rule,
  \be
  \psi({\bm\tau})\to\left|c{\bm\tau}+d\right|^k\psi({\bm\tau}),
  \ee
  },
  \be
  \psi({\bm\tau})\to\left|c{\bm\tau}+d\right|\psi({\bm\tau}),
  \label{i1}
  \ee
  under the modular transformation,
  \be
  {\bm \tau}\to{\bm\tau}'=\frac{a{\bm\tau}+b}{c{\bm\tau}+d},\hspace{1cm}\left(\begin{array}{cc}a&b\\c&d\end{array}\right)\in
  {\rm SL}(2,\mathbb{Z}).
  \label{i2}
  \ee
  Similar results are obtained for
  K\"{a}hler quantization of $H^2(\mathbb{T}^2,\mathbb{R})$.
  The weight of the wave function is characteristic to the operator
  ordering. By operator ordering one means the ordering of the
  operators e.g. $\hat x$ and $\hat p$, corresponding to the coordinate $x$ and the conjugate
  momentum $p$ in e.g. the expression
  $xp$. If one uses the standard convention $xp\to\frac{1}{2}[\hat x,\hat p]_+$
  where $[\hat x,\hat p]_-=i\hbar$ is assumed, then the wave
  function in the K\"{a}hler quantization of
  $H^*(\mathbb{T}^2,\mathbb{R})$ is a quasi-modular
  form of weight one as is given in Eq.(\ref{i1}). But if for example
  one uses the unconventional operator ordering $xp\to\hat x\hat p$
  while assuming $[\hat x,\hat p]_-=i\hbar$, it can be shown that the
  wave function is a quasi-modular form of weight zero, i.e. it is
  invariant under the modular group SL$(2,\mathbb{Z})$.

  The organization of the paper is as follows. In section
  \ref{torus} our notation, conventions and some basic calculations are given. In section \ref{H1}, K\"{a}hler quantization of
  $H^{1}(\mathbb{T}^2,\mathbb{R})$ is studied and the moduli-independence conditions are solved for the quantum wave
  function.  In section {\ref{H2}}, K\"{a}hler quantization of
  $H^{2}(\mathbb{T}^2,\mathbb{R})$ is studied.  In section \ref{fermion} we show that the cohomology $H^1(\mathbb{T}^2,\mathbb{R})$
  corresponds to the fermionic $\sigma$-model $\mathbb{R}\to{\cal
  P}_2$. Furthermore it is shown that ${\cal P}_2$ is a
  noncommutative space.
  In section \ref{ordering} the dependence of the
  weight of the wave function on the operator ordering is examined in the
  unconventional ordering $xp\to\hat x\hat p$.
 \section{Preliminaries}\label{torus}
  In this section we derive necessary equations used in the subsequent sections and give our
  notations and conventions.

\begin{figure}[t]
 \centerline{\epsfxsize=3in\epsffile{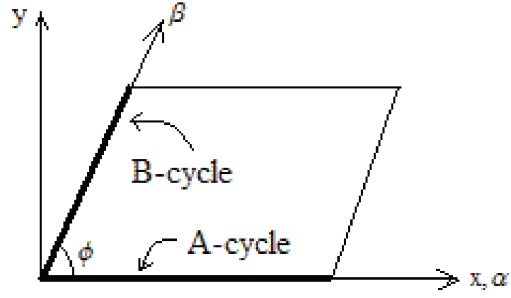}} \caption{$\mathbb{T}^2$: the length of $A$-cycle is 1 and the length of $B$-cycle is $\tau$.}
 \label{fig}
 \end{figure}

  Take a complex plane $\mathbb{C}$ and define a lattice
  $L(\omega_1,\omega_2)=\{\omega_1m+\omega_2n|m,n\in\mathbb{Z}\}$
  where $\omega_1$ and $\omega_2$ are nonvanishing complex numbers
  such that $\omega_2/\omega_1\not\in\mathbb{R}$. A two-dimensional torus
  $\mathbb{T}^2\cong\mathbb{C}/L(\omega_1,\omega_2)$ is obtained by identifying the
  points $z_1$, $z_2\in\mathbb{C}$ such that
  $z_1-z_2=\omega_1m+\omega_2n$ for some $m$, $n\in\mathbb{Z}$. The complex structure on
  $\mathbb{T}^2$ is defined by the pair of complex numbers $(\omega_1,\omega_2)$
  modulo a constant factor and PSL$(2,\mathbb{Z})$ \cite{Koblitz}. Consequently,
  one can take 1 and the modular parameter ${\bm \tau}=\omega_2/\omega_1$
  to be the generators of the lattice. The complex structure of
  $\mathbb{T}^2$ is thus specified by ${\bm\tau}$. Furthermore, ${\bm \tau}$
  and ${\bm \tau}'=(a{\bm \tau}+b)/(c{\bm \tau}+d)$ define the same complex structure
  if,
  \be
  \left(\begin{array}{cc}a&b\\c&d\end{array}\right)\in{\rm
  SL}(2,\mathbb{Z}).
  \label{t1}
  \ee
 To any point $(x,y)\in L(1,{\bm \tau})$ we assign coordinates $(\alpha,\beta)\in \mathbb{T}^2$ (see
 Fig.\ref{fig}),
 \be
 \alpha=x-y\cot\phi,\hspace{1cm}\beta=\frac{y}{\tau\sin\phi},
 \ee
 in which $\tau$ and $\phi$ are moduli defined in terms of the modular parameter
 ${\bm {\tau}}=\tau e^{i\phi}$.
 By definition, ,
 \be
 \begin{array}{ccc}\int_Ad\alpha=1,&&\int_B d\alpha=0,\\\\
 \int_A d\beta=0,&&\int_B d\beta=1.
 \end{array}
 \label{t2}
 \ee
 In real polarization, a one-form $\Omega\in H^{1}(\mathbb{T}^2,\mathbb{R})$ is
 given in terms of its periods,
 \be
 h_r=\int_A\Omega,\hspace{1cm}h_i=-\int_B\Omega.
 \label{t3}
 \ee
 Thus,
 \be
 \Omega=h_rd\alpha-h_id\beta.
 \label{t3-1}
 \ee
 Under an infinitesimal moduli variation $\delta\tau$ and
 $\delta\phi$,
 \bea
 \delta\alpha&=&\frac{\tau\beta}{\sin\phi}\delta\phi,\nn\\
 \delta\beta&=&-\left(\frac{1}{\tau}\delta\tau+\cot\phi\delta\phi\right)\beta.
 \label{t4}
 \eea
 Since $\Omega$ is independent of moduli, i.e. $\delta\Omega=0$, one
 obtains,
 \bea
 \delta h_r&=&0,\nn\\
 \delta
 h_i&=&\frac{h_r\tau}{\sin\phi}\delta\phi+h_i\left(\frac{\delta\tau}{\tau}+\cot\phi\delta\phi\right).
 \label{t4-1}
 \eea
 On the other hand, under modular transformation (\ref{i2}),
 \bea
 \tau\cos\phi&\to&\tau'\cos\phi'=\frac{\tau\cos\phi}{\left|c{\bm\tau}+d\right|^2}+\frac{ac\tau^2+2bc\tau\cos\phi+bd}{\left|c{\bm\tau}+d\right|^2},\nn\\
 \tau\sin\phi&\to&\tau'\sin\phi'=\frac{\tau\sin\phi}{\left|c{\bm\tau}+d\right|^2},
 \label{t5}
 \eea
 thus,
 \bea
 \alpha&\to&\alpha'=\alpha-\beta\left(ac\tau^2+2bc\tau\cos\phi+bd\right),\nn\\
 \beta&\to&\beta'=\left|c{\bm\tau}+d\right|^2\beta.
 \label{t6}
 \eea
 $\Omega$ is invariant under modular transformation. Thus under modular transformation,
 \bea
  h_r&\to&h_r'=h_r,\nn\\
  h_i&\to&h_i'=\frac{h_i}{\left|c{\bm\tau}+d\right|^2}-h_r\frac{ac\tau^2+2bc\tau\cos\phi+bd}{\left|c{\bm\tau}+d\right|^2}.
  \label{t6-1}
  \eea
  A two form $\Xi\in H^{2}(\mathbb{T}^2,\mathbb{R})$ in real polarization is given
  by $\Xi=\xi d\alpha\wedge d\beta$. Assuming invariance of $\Xi$ under infinitesimal moduli variation, it is easy to show
  that,
  \be
  \delta\xi=\left(\frac{\delta
  \tau}{\tau}+\cot\phi\delta\phi\right)\xi.
  \label{t7}
  \ee
  Furthermore under modular transformation,
  \be
  \xi\to\xi'=\frac{\xi}{\left|c{\bm\tau}+d\right|^2}.
  \label{t7-1}
  \ee
 \section{K\"{a}hler quantization of
  $H^{1}(\mathbb{T}^2,\mathbb{R})$}\label{H1}
  Using Eq.(\ref{t4-1}), the classical generators of translation along moduli directions can
  be defined as follows,
  \bea
  H_\tau&=&\frac{1}{\tau}h_i\pi_i,\nn\\
  H_\phi&=&\left(\frac{h_r\tau}{\sin\phi}+h_i\cot\phi\right)\pi_i,
  \label{H1-1}
  \eea
  in which $\pi_i$ is the momentum conjugate to $h_i$ satisfying the
  Poisson algebra $\{h_i,\pi_i\}=1$. For quantization, one replaces
  the classical fields with the corresponding operators and uses the
  standard commutation relation,
  \be
  [\hat h_i,\hat \pi_i]=i.
  \label{H1-2}
  \ee
  Henceforth by e.g. $h_i$ we mean the operator $\hat h_i$.

  Using  the standard operator ordering the quantum mechanical generators
  of translation along moduli directions are defined as,
 \bea
  H_\tau&=&\frac{1}{\tau}\frac{[h_i,\pi_i]_+}{2}=-\frac{i}{\tau}\left(h_i\frac{\partial}{\partial h_i}+\frac{1}{2}\right),\nn\\
  H_\phi&=&\frac{h_r\tau}{\sin\phi}\pi_i+\frac{[h_i,\pi_i]_+}{2}\cot\phi=-i\frac{h_r\tau}{\sin\phi}\frac{\partial}{\partial
  h_i}-i  \cot\phi\left(h_i\frac{\partial}{\partial h_i}+\frac{1}{2}\right).
  \label{H1-3}
  \eea
  The background independent wave function $\psi$ is defined by the
  following relations \cite{Kahler},
  \bea
  i\frac{\partial}{\partial \tau}\psi&=&H_\tau\psi,\nn\\
  i\frac{\partial}{\partial \phi}\psi&=&H_\phi\psi.
  \label{H1-4}
  \eea
  Solving for $\psi$ one obtains,
  \be
  \psi=\left\{\begin{array}{lll}
  \frac{1}{\sqrt{
  h_i}}f\left(\frac{h_i}{\tau\sin\phi}\right),&&h_r=0,\\\\
  \frac{c_1\sqrt{\tau\sin\phi}}{h_i+h_r\tau\cos\phi}+\frac{c_2}{\sqrt{\tau\sin\phi}},&&h_r\neq
  0,
  \end{array}\right.
  \label{H1-5}
  \ee
  in which $f$ is an arbitrary ${\cal C}^1$ function and $c_1$ and
  $c_2$ are two arbitrary constants. Using Eqs.(\ref{t6}) and
  (\ref{t6-1}) one can simply show that the wave function $\psi$ is a
  quasi-modular form of weight one. See Eq.(\ref{i1}).
  \subsection{K\"{a}hler quantization of
  $H^{2}(\mathbb{T}^2,\mathbb{R})$}\label{H2}
  Using Eq.(\ref{t7}) one can obtain the quantum mechanical generators
 of translation along moduli directions,
  \bea
  H_{\xi\tau}&=&-\frac{i}{\tau}\left(\xi\frac{\partial}{\partial
  \xi}+\frac{1}{2}\right),\nn\\
  H_{\xi\phi}&=&-i\cot\phi\left(\xi\frac{\partial}{\partial
  \xi}+\frac{1}{2}\right).
  \label{H2-1}
  \eea
  The corresponding quantum background independent wave function
  satisfies,
 \bea
  i\frac{\partial}{\partial \tau}\psi_\xi&=&H_{\xi\tau}\psi_\xi,\nn\\
  i\frac{\partial}{\partial \phi}\psi_\xi&=&H_{\xi\phi}\psi_\xi.
  \label{H2-2}
  \eea
  Thus,
  \be
  \psi_\xi=\frac{1}{\sqrt{
  \xi}}f_\xi\left(\frac{\xi}{\tau\sin\phi}\right),
  \label{H2-3}
  \ee
  where $f_\xi$ is an arbitrary ${\cal C}^1$ function. Using
  Eqs.(\ref{t6}) and (\ref{t7-1}), one can  show that $\psi_\xi$ is a
  quasi-modular form of weight one.
 \section{Fermions on a line}\label{fermion}
  The cohomology $H^1(\mathbb{T}^2,\mathbb{R})$ has the structure of a phase space. Similar to the cohomology $H^3({\rm CY}_3,\mathbb{R})$ which
  describes the classical solutions for the 7 dimensional action \cite{Verlinde,Kahler},
  \be
  S=\frac{1}{2}\int_{{\rm CY}_3\times\mathbb{R}}\gamma\wedge
  d\gamma,
  \label{f1}
  \ee
  $H^1(\mathbb{T}^2,\mathbb{R})$ describes the classical solutions
  for the 3 dimensional action,
  \be
  S\sim\int_{\mathbb{T}^2\times\mathbb{R}}\eta\wedge d\eta,
  \label{f2}
  \ee
 in which $\eta$ is a real one-form on $\mathbb{T}^2\times\mathbb{R}$.
 \be
 \eta=\Omega+\rho dt
 \label{f3}
 \ee
 where $\Omega$ is a real one form along $\mathbb{T}^2$, $\rho$ is a zero-form and $t$ parameterizes $\mathbb{R}$.
 This claim can be verified by noting that the action (\ref{f2}) is equivalent to,
 \be
 S\sim\int_{\mathbb{R}}\int_{\mathbb{T}^2}\left(-\Omega\wedge\partial_t\Omega-2\rho
 d_{\mathbb{T}^2}\Omega\right).
 \label{f4}
 \ee
 Obviously, the equation of motion corresponding to $\rho$ implies that
 \be
 d_{\mathbb{T}^2}\Omega=0.
 \label{f5}
 \ee
  Therefore, $\Omega$ is a real closed one form which has been the subject
 of study in section \ref{H1}.

 Using Eqs.(\ref{t3}) and (\ref{t3-1}), the action (\ref{f4}) can be written as
 follows,
 \be
 S\sim\int_{\mathbb{R}}\left(h_r\partial_th_i-h_i\partial_th_r\right)dt.
 \label{f6}
 \ee
 Thus defining the {\em spinor} $h$,
 \be
 h=\left(\begin{array}{c}h_r\\h_i\end{array}\right),
 \label{f7}
 \ee
 and the gamma matrix,
 \be
 \gamma^0=\left(\begin{array}{ccc}0&&i\\-i&&0\end{array}\right),
 \label{f8}
 \ee
 it can be verified that action (\ref{f6}) describes free fermions
 $h$ on a real line $\mathbb{R}$,
 \be
 S\sim\int\left(-ih^t\gamma^0\partial_t h\right)dt.
 \label{f9}
 \ee
 It should be noted that the spinor field $h$ is subject to the
 identification (\ref{t6-1}) under modular group SL($2,\mathbb{Z}$).
 Consequently, the action (\ref{f9}) describes the fermionic
 $\sigma$-model $\mathbb{R}\to{\cal P}_2$ noted in section
 \ref{Int}.

 ${\cal P}_2$ is a noncommutative space since it is described by a first
 order Lagrangian. Indeed, using Eq.(\ref{f6}) one obtains,
 \bea
 \pi_r&=&\frac{\delta S}{\delta \partial_th_r}=-bh_i,\nn\\
 \pi_i&=&\frac{\delta S}{\delta \partial_th_i}=bh_r,
 \label{f10}
 \eea
 in which $b$ is the unspecified constant coefficient in the definition of the
 action (\ref{f6}). Therefore, the action (\ref{f6}) is describing a
 constrained system of second class \cite{Dirac}. The Dirac bracket
 algebra implies that,
 \be
 \{h_i,h_r\}_{\rm DB}=\frac{1}{2b},
 \label{f11}
 \ee
 though the corresponding Poisson bracket vanishes. Consequently the target
 space ${\cal P}_2$ is equipped with a noncommutative structure. In
 \cite{Kahler}, in a similar situation, the coefficient of the
 action (\ref{f1}) is determined by requiring that quantization of $H^3({\rm
 CY}_3,\mathbb{R})$ gives exactly the holomorphic anomaly equation
 in the topological string model \cite{BCOV}. Here, $b$ remains
 undetermined.

  \section{Dependence of the weight of quasi-modular form on operator
  ordering}\label{ordering}
  In this section we examine the dependence of the
  weight of quasi-modular form on the operator ordering by considering the
  unconventional ordering $xp\to\hat x\hat p$.
  Under this ordering, the quantum mechanical generators
  of translation along moduli directions $H_\tau$ and $H_\phi$ in the
  K\"{a}hler  quantization of $H^1(\mathbb{T}^2,\mathbb{R})$ that can be obtained from
  Eq.(\ref{H1-1}) are,
  \bea
  H_\tau&=&-\frac{i}{\tau}h_i\frac{\partial}{\partial h_i},\nn\\
  H_\phi&=&-i\frac{h_r\tau}{\sin\phi}\frac{\partial}{\partial
  h_i}-i  \cot\phi h_i\frac{\partial}{\partial h_i}.
  \label{o1}
  \eea
 The solution to the corresponding wave-equation is,
  \be
  \psi=\left\{\begin{array}{lll}
   f\left(\frac{h_i}{\tau\sin\phi}\right),&&h_r=0,\\\\
  \frac{h_r\tau\sin\phi}{h_i+h_r\tau\cos\phi}c_1+c_2,&&h_r\neq0,
  \end{array}\right.
  \label{o2}
  \ee
   This can be easily shown to be invariant under the modular
   transformation (\ref{t5}) and (\ref{t6-1}). Thus under the
   unconventional operator ordering $xp\to\hat x\hat p$, the background independent wave function
   $\psi$ in real polarization  becomes a quasi-modular form of weight zero.

   A similar result can be obtained in the K\"{a}hler
  quantization of $H^2(\mathbb{T}^2,\mathbb{R})$.
  Here,
  \bea
  H_{\xi\tau}&=&-\frac{i}{\tau}\xi\frac{\partial}{\partial
  \xi},\nn\\
  H_{\xi\phi}&=&-i\cot\phi\xi\frac{\partial}{\partial
  \xi}.
  \label{o3}
  \eea
 and thus,
 \be
  \psi_\xi=f_\xi\left(\frac{\xi}{\tau\sin\phi}\right),
  \label{o4}
  \ee
  which is invariant under the modular transformation.
 \section*{Summary}
  In the K\"{a}hler  quantization of $H^3({\rm CY}_3,\mathbb{R})$, the genus $g$ partition
 function  is shown to be an almost modular form
   \cite{Aganagic}. Here we showed that in the K\"{a}hler
  quantization of $H^*(\mathbb{T}^2,\mathbb{R})$, the background independent wave function
   in real polarization is a quasi-modular form of weight one in the
   standard operator ordering $xp\to\frac{1}{2}[\hat x,\hat p]_+$, while it is a quasi-modular form of weight zero in
   the unconventional operator ordering $xp\to\hat x\hat p$.
 \section*{Acknowledgement}
 The financial support of Isfahan University of Technology (IUT) is
 acknowledged.
 
\end{document}